\documentstyle[12pt,psfig]{article}
\begin{document}
\thispagestyle{empty}
\noindent\hbox to\hsize{\hfill  November 1997}\\
\noindent\hbox to\hsize{\hfill Brown-HET-1105}\\
\noindent\hbox to\hsize{\hfill Brown-TA-553}\\
\noindent\hbox to\hsize{\hfill  ULG-PNT-97-2-JRC}\vskip 0.6in
\vskip 0.3in
\begin{center}
{\bf SOFT POMERON INTERCEPT}
\footnote{presented by K. Kang 
at the International Conference (VIIth Blois Workshop) on Elastic and 
Diffractive Scattering, {\it Recent Advances in Hadron Physics}, 
Ewha Womans University, Seoul, Korea, 10 - 14 June 1996}\\ 
\vskip 0.2in
Jean-Ren\'e Cudell\\{\small 
Inst. de Physique, U. de Li\`ege,
B\^at. B-5, Sart Tilman, B4000 Li\`ege, Belgium\\
\baselineskip 5pt cudell@gw.unipc.ulg.ac.be}\\
%~\\
 Kyungsik Kang\footnote{supported in part by the 
US DOE contract DE-FG02-91ER 40688-Task A.}
\\
{\small Physics Department, Brown University,
Providence, RI 02912, U.S.A.\\kang@het.brown.edu}\\
%~\\
%and\\~\\
 Sung Ku Kim\footnote{supported in part by the Korea Science and Engineering 
Foundation through the Center for Theoretical Physics, Seoul National
University, and by the 
Ministry of Education through contract BSRI-95-2427.}\\
{\small Dept. of Physics, Ewha Women's University, Seoul 120-750, Korea\\
skkim@mm.ewha.ac.kr} \\
~\\
{(presented by Kyungsik Kang)}
\vskip 0.3in

{\bf Abstract}
\end{center}
\begin{quote}
We show that simple Regge pole 
fits to the imaginary and the real part of the elastic amplitude in $pp$ and $\bar p p$ collisions give a value of $1.096^{+0.012}_{-0.009}$ for the pomeron intercept. Preliminary results of the new global fits to all hadron total cross sections as well as
to the proton structure function at low x and low $q^2$ are briefly mentioned.

\end{quote}\newpage

The simplest object responsible for the rising behavior of total cross sections would 
be a Regge trajectory, the pomeron, with an intercept somewhat greater than 1. This object has been hypothesized a long time ago \cite{Collins}, 
and presumably
arises from the gauge sector of QCD. Its intercept and slope are then
fundamental numbers characterizing the pure gauge sector of Yang-Mills SU(3).
After the pioneering work of UA8 \cite{UA8}, there has been a renewed interest in the pomeron
after the observation of rapidity gaps in deep inelastic $ep$ scattering
at HERA
\cite{gaps} 
and the suggestion 
that such cross sections might be used for the detection of new physics
\cite{Bj}. The deep-inelastic diffractive scattering experiment can offer a unique opportunity to study the soft process with a hard virtual photon probe and can shed light on the formulation of Reggeon field theory that can explain both the soft and hard scattering and hopefully lead to a more complete understanding of QCD.
One can view the emergence of gaps as resulting from the emission of 
a pomeron by the colliding proton (which then remains in a color-singlet
state) followed by a pomeron-photon collision. 
The validity of this picture, as
well as the measurement of the pomeron intercept and structure function
is thus a central issue for HERA, and will no doubt have a bearing on the 
extrapolation to the physics at the LHC energies.
Hence it will be very important to have a precise determination of
%o re-evaluate 
the pomeron intercept. We have re-evaluated the pomeron intercept as carefully as possible from the simple-pole model fits to the total cross sections 
and estimated
the errors on the various parameters \cite{CKK}. 

The information on the soft pomeron intercept comes from 
the high energy behavior of the total cross sections. Because of the presence of sub-leading Regge poles at low energy and because of the unitarity effect
due to multipomeron exchanges at high energy, one must not only determine the best parameterization but also the range of validity of the model. It is clear that a simple-minded $\chi^2$ test cannot be sufficient, primarily because 
the cross section data contain many points that are inconsistent with 
their neighboring points. We therefore must invent a reasonable method to filter the data sets independently of any underlying theoretical model or prejudice. A reasonable criterion is that a given data
point should not deviate by more than 1 or 2$\sigma$ from the average of all data in a bin of $\pm$1 GeV centered around it and yet 
the central values of the parameters and their errors should not depend too sensitively on the filtering itself. 
%\section{The definition of the errors}

In the following, we use two strategies to evaluate the best 
central values and their
errors. The first is to use all the data available \cite{dataset, Kangdata}. This gives us
the central values. However, some of the data points are incompatible at the $2\sigma$
level or more. This means that the value of $\chi^2$ will be artificially inflated primarily due to inclusion of those few incompatible data points. 
In fact the best fits \cite{Kangfit, PDG} that
one can produce have typically a $\chi^2/d.o.f.$ of 1.3 or more.
We then use the data sets that are filtered by using the proposed criterion, which will give us stable central values of the parameters and their 
uncertainties. 
We give in Table 1 the number of data points that are kept before and after filtering. The full data sets are available at http://nuclth02.phys.ulg.ac.be/Data.html.
\begin{quote}{\small
\begin{tabular}{|c|c|c|c|c|}\hline
data set&$\sigma_{tot}^{pp}$ (mb) &$\sigma_{tot}^{\bar pp}$ (mb)&$\rho^{pp}$&$
\rho^{\bar pp}$ \\ \hline
P.D.G. \cite{dataset}&94&28&-&-\\
P.D.G. \cite{dataset} - $2\sigma$ &84&28&-&-\\
P.D.G. \cite{dataset} - $1\sigma$ &65&20&-&-\\
Ref.~\cite{Kangdata}&66&29&41&13\\
Ref.~\cite{Kangdata} - $2\sigma$&60&28&38&13\\
Ref.~\cite{Kangdata}- $1\sigma$&53&19&31&13\\\hline
\end{tabular}}
{}~\\
\end{quote}\begin{quote}
{\small Table 1: The number of points kept after data selection, for \\
\hbox{$\sqrt{s}>10$~GeV.}}
\end{quote}
Note that the $2\sigma$ selection includes both CDF and E710 points, whereas
the $1\sigma$ one rejects them both. This procedure is not unlike the one
followed by UA4/2 in \cite{UA4}.
As the value
of $\chi^2$ - $\chi^2_{min}$ is distributed as a $\chi^2$ with N parameters of the model, the $\Delta \chi^2$ corresponding to $70 \%$ confidence level(C.L.) is 6.06 \cite{minuit, statistics}
in the DL case with 5 parameters. The errors we quote in this paper are to this
$\chi^2$-interval, to be contrasted to those quoted in the Particle Data Group(PDG)
\cite{PDG} who
simply renormalized the $\chi^2$ to $\chi^2 /d.o.f. = 1$ and let the new $\chi^2$
change from the minimum by one unit.

Donnachie and Landshoff (DL) have proposed\cite{DL}
to fit the $pp$ and $\bar p p$ cross section
using a minimal number of trajectories: the leading meson trajectories of the
degenerate $a/f$ ($C=+1$) and $\rho/\omega$ ($C=-1$), plus the pomeron trajectory. They fit data
for $s>100$ GeV$^2$, as lower trajectories would then contribute 
less than 1\%, which
is less than the errors on the data. The result of their fit \cite{DLfit}
is a pomeron 
intercept of 1.0808, for which they did not quote a $\chi^2$ or error bars and said that the $\chi^2$ was very flat near the minimum.
We show in Table 2 our results for such a fit. We use the usual definition 
\begin{equation}
\chi^2=\sum_i \left({d_i-s_i^{-1} {\cal I}m{\cal A}(s_i,0))\over e_i}\right)^2
\label{chi2}
\end{equation}
with $d_i\pm e_i$ for the measured  $pp$ or $\bar p p$ total cross section 
at energy $\sqrt{s_i}$, and
\begin{equation}
{\cal I}m{\cal A}(s,0)=C_- s^{\alpha_m} + C_+ s^{\alpha_m} +C_P s^{\alpha_P}
\label{amplitude}
\end{equation}
where $C_-$ flips sign when going from $pp$ to $\bar p p$. 
We use the same data set as DL \cite{dataset} to determine the central value,
and use the selected data sets at the 1- or 2-$\sigma$ level to determine the errors. We see that the fit to all 
data gives a totally unacceptable value, $\chi^2=410$ for 135 data points with 5 parameters, corresponding to a C.L. of
$2\times 10^{-36}$ !
There are two possible outcomes to such a high value of the $\chi^2$:
either the model is to
be rejected, or some of the data are wrong. We adopt the second view and try
to eliminate those points.  
\begin{figure}
\begin{center}
%{\small \begin{tabular}{|c|c|c|c|}\hline
{}~\vglue-12pt
{\small \begin{tabular}{|c|c|c|c|}\hline
parameter&  all data&filtered data (2$\sigma$)&filtered data
(1$\sigma$)\\\hline
$\chi^2	$	&410.8	&80.3 &32.4\\
$\chi^2$ per d.o.f.	&3.16 &0.62&0.25\\\hline
pomeron intercept-1
&$0.0912^{+0.0077}_{-0.0070}$&$0.0887^{+0.0079}_{-0.0071}$
&$0.0863^{+0.0096}_{-0.0084}$\\
pomeron coupling (mb)&$19.3^{+1.5}_{-1.7}$& $19.8^{+1.6}_{-1.7}$ &
$20.4^{+1.8}_{-2.1}$ \\\hline
$\rho/\omega/a/f$ intercept-1
&$-0.382^{+0.065}_{-0.071}$&$-0.373^{+0.067}_{-0.073}$
&$-0.398^{+0.083}_{-0.090}$\\
$\rho$/$\omega$ coupling $C_-(pp)$ (mb)
&$-13.2^{+4.1}_{-6.5}$&$-13.3^{+4.3}_{-6.9}$&$-15.3^{+5.7}_{-10.1}$\\
$a/f$ coupling (mb)		&
$69^{+20}_{-13}$&$62^{+19}_{-12}$&$67^{+25}_{-16}$\\\hline
\end{tabular}}~\\
\end{center}
\begin{quote}
{\small Table 2: Simple pole fits to total $pp$ and $\bar p p$ cross sections,
assuming
degenerate $C=+1$ and $C=-1$ exchanges. }
\end{quote}
\end{figure}
Table~2 shows that  the central values
and their errors are indeed stable against
%do not depend too much on 
the filtering itself, while
filtering the data does improve the value of
$\chi^2_{min}$ drastically so that the model becomes perfectly acceptable. Also
%this is indeed the case for our fit. we see
note that the pomeron intercept is determined to be about 1.090, and that
it could be as high as 1.096.
%but does not affect its variation around the minimum too much.
We think the stability of the parameter values is a more
important than the value of $\chi^2_{min}$ itself.

As for the energy range of validity of the model, we adopt two basic requirements: that the $\chi^2/d.o.f.$ be of the order of 1,
and that the determination of the intercept be stable. We show in Fig.~1
the result of varying the energy range. Clearly, the lower trajectories seem
to matter for $\sqrt{s}_{min}<10$ GeV, whereas the upper energy does not seem
to
modify the results (in other words, there is no significant 
sign of the onset of unitarisation up to the Tevatron energy). Hence
we adopt $\sqrt{s}_{min}=10$ GeV as the lowest energy at which the model is
correct. This happens to be the point at which $\chi^2_{min}$ is lowest.
This dependence on the lower energy cut explains why both
the PDG \cite{PDG} and Bueno and Velasco \cite{BV}
obtain an wrong value for the intercept.
%, much lower than ours.

%{}~\\
\hbox{
\psfig{figure=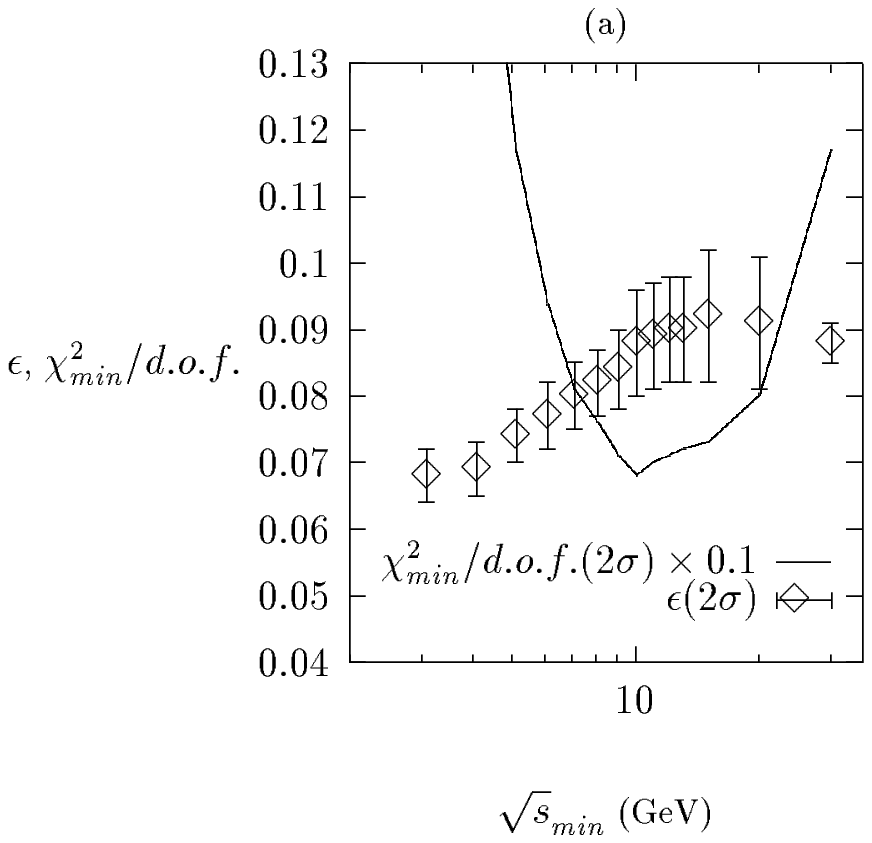,bbllx=2.5cm,bblly=15cm,bburx=12cm,bbury=23cm,width=6cm}
\psfig{figure=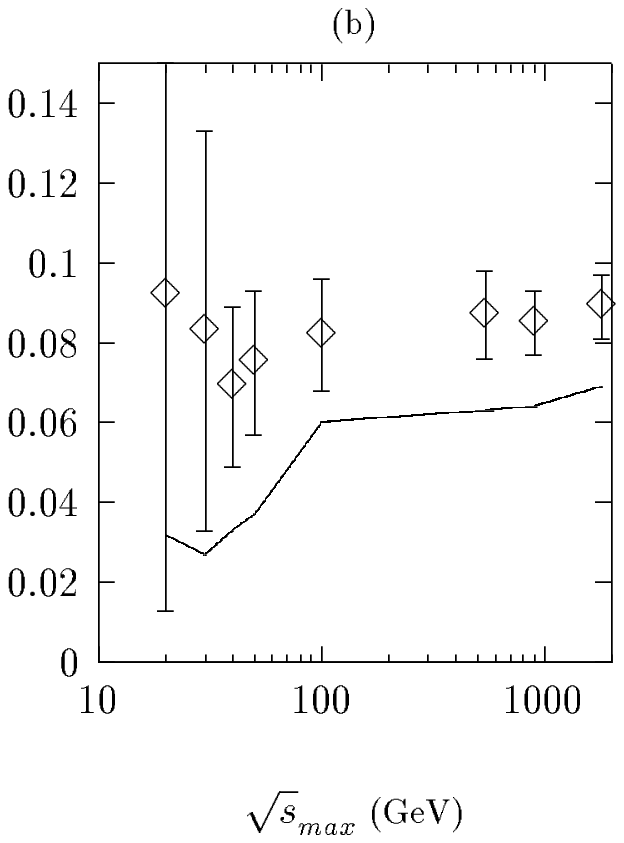,bbllx=2.5cm,bblly=15cm,bburx=12cm,bbury=23cm,width=6cm}
}
\begin{quote}
{\small Figure 1: DL intercept-1 as a function of the lower (a) and upper (b)
energy cuts
on the data. The curve shows the $\chi^2/dof.$ for data filtered at
the $2\sigma$ level.
}
\end{quote}

One must wonder if it is possible to get a better determination of the soft
pomeron intercept by using more data. 
The PDG \cite{PDG} obtained very narrow determinations
of the pomeron intercept from the other hadronic reactions. We believe
that their conclusions were wrong, and illustrate this in the case of
the $\pi^{\pm} p$ total cross sections, for which they used
$\sqrt{s}_{min}\approx 4$ GeV and obtained an intercept of $1.079\pm 0.003$.
%We show in Fig.~2 
Our results for the data set filtered at the
$2\sigma$ level (92 points), and for $\sqrt{s}_{min}=4$ GeV, give
%we obtain
$\alpha_P=1.115^{+0.030}_{-0.023}$, which is consistent with the one we got
from the analysis of $pp$ and $\bar p p$ total cross sections. 
%There is no significant change if
%we modify the lower energy cutoff on the data. We show in 
%Fig.~2(b) shows our
%best fit together with that of the PDG \cite{PDG}. 
Although according to their previous estimate our central
value for the intercept was 10 standard deviations from theirs, it turns out
%we see 
that our best fit and that of the PDG \cite{PDG} are indistinguishable. 
Hence we believe that both their
standard deviations and their central values were wrong. These conclusions are
not affected by the use of the full data set. A recent reanalyses has confirmed this
point of view \cite{PDG97} and our determination of the intercept is now in agreement with the new analysis of the PDG. 
%instead of the one filtered at the $2\sigma$ level.
%{}~\\
%\hbox{
%\psfig{figure=pir.ps,bbllx=2.5cm,bblly=7.5cm,bburx=12cm,bbury=23cm,width=6cm}
%\psfig{figure=pif.ps,bbllx=2.5cm,bblly=7.5cm,bburx=12cm,bbury=23cm,width=6cm}
%}
%\begin{quote}
%{\small Figure 2: (a) shows the pomeron intercept from $\pi p$ data as the %lower energy cut on the data is changed; (b) shows our
%best fit to the data set filtered at the $2\sigma$ level together with that of %the Particle Data
%Group.
%}
%\label{pirange}
%\end{quote}
%Note that this intercept is consistent with the one we got
%from the analysis of $pp$ and $\bar p p$ total cross sections.
%The conclusion from this exercise is that the errors from the low-energy
%hadronic data are large, especially if we use a low-energy cut-off of the
%order of 10. Hence we want to limit ourselves to
%$pp$ and $\bar p p$ amplitudes at $t=0$.

One more
piece of $pp$ and $\bar p p$ data can be used however:
the knowledge of the intercept is sufficient to determine the value
of the real part of the amplitude, using crossing symmetry, and hence
the measurements of the $\rho$ parameter provide an extra constraint.
We use the data
collected in Ref.~\cite{Kangdata}, and obtain a somewhat worse fit,
as shown in Table~3, even
when filtering data at the $1$ or $2\sigma$ level. For the 2$\sigma$ filtering
of the data, the C.L. goes from 99.4\% to 36\%.
We show the curves corresponding to the second column of Table~3 in Fig.~2 with
dotted lines. Whether one
should worry about the change of $\chi^2$ and the change of central value for the parameters, is a matter of taste. But it is this small change of
central values, combined with the effect of too low an energy cut, that lead
Bueno and Velasco \cite{BV} to conclude that simple-pole parametrisations were
disfavored.
\begin{center}
{\small \begin{tabular}{|c|c|c|c|}\hline
parameter&  all data& filtered data ($2 \sigma$)&filtered data ($1\sigma$)\\
\hline
$\chi^2	$		& 561.3	&168.3 &94.9\\
$\chi^2$ per d.o.f.	& 3.28	&1.07&0.77\\\hline
pomeron intercept--1	&$0.0840\pm
0.0050$&$0.0817^{+0.0055}_{-0.0053}$&$0.0804^{+0.0064}_{-0.0061}$\\
pomeron coupling (mb)	& $20.8 \pm 1.1$&$21.4\pm 1.1$& $21.8\pm 1.3$\\\hline
$\rho/\omega/a/f$
intercept--1&$-0.408^{+0.032}_{-0.033}$&$-0.421^{+0.034}_{-0.036}$&
$-0.431^{+0.037}_{-0.040}$\\
$\rho$/$\omega$ coupling $C_-(pp)$ (mb)&$-14.0 ^{+2.6}_{-3.3}$&$-16.5
^{+3.3}_{-4.2}$&$-17.7^{+3.7}_{-4.9}$\\
$a/f$ coupling (mb)		& $67.0 ^{+7.6}_{-6.7}$&$66.6 ^{+8.3}_{-7.2}$&$67.6
^{+9.0}_{-7.8}$\\\hline
\end{tabular}}~\\
\end{center}\begin{quote}
{\small Table 3: Simple pole fit to total $pp$ and $\bar p p$ cross sections,
and to the $\rho$
parameter, assuming
degenerate $C=+1$ and $C=-1$ meson exchanges.}
\end{quote}

Before concluding on the best value of the intercept, we need to examine
the influence of low energy cut on the determination of the intercept.
Although the energy cut $\sqrt{s}_{min}$ eliminates sub-leading meson
trajectories, there is still an ambiguity in the treatment of the leading
meson trajectories. In fact, a slightly different
treatment to that of DL leads to a better $\chi^2$ and to more
stable parameters.
Indeed, there is neither theoretical nor experimental reason to assume that the $\rho$, $\omega$, $f$ and
$a$ trajectories are degenerate. On the other hand, the data are not constraining enough to determine the
effective intercepts of the four
meson trajectories together with the pomeron intercept. We adopt
an intermediate approach, which is to assume the exchange of separate + and $-$
trajectories with independent intercepts:
\begin{equation}
Im{\cal A}(s,0)=C_- s^{\alpha_-} + C_+ s^{\alpha_+} +C_P s^{\alpha_P}
\label{2traj}
\end{equation}
The resulting numbers are shown in Table~4, and are plotted in
Fig.~2 with plain lines.  The $\chi^2$ is smaller
and the parameters are more stable than in the previous case. 
The bounds on the soft pomeron intercept
hardly depend on the criterion used to filter the data, and
intercepts as large as 1.108 are allowed.
In order to better understand the treatment of the errors, we give in Table~5
the result of a fit to the data of Ref.~\cite{Kangdata}. Our results are very stable, especially those for the pomeron intercept, independently of the data filtering.
 
%{}~\\
\hbox{
\psfig{figure=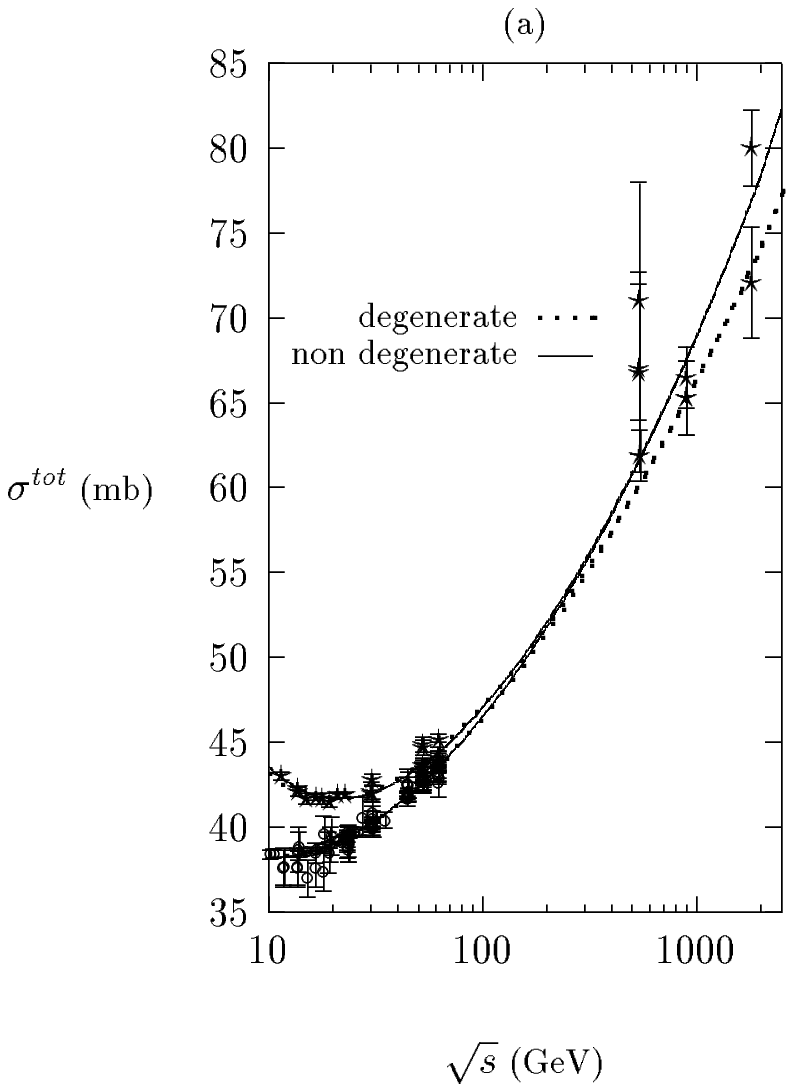,bbllx=2.5cm,bblly=12cm,bburx=12cm,bbury=23cm,width=6cm}
\psfig{figure=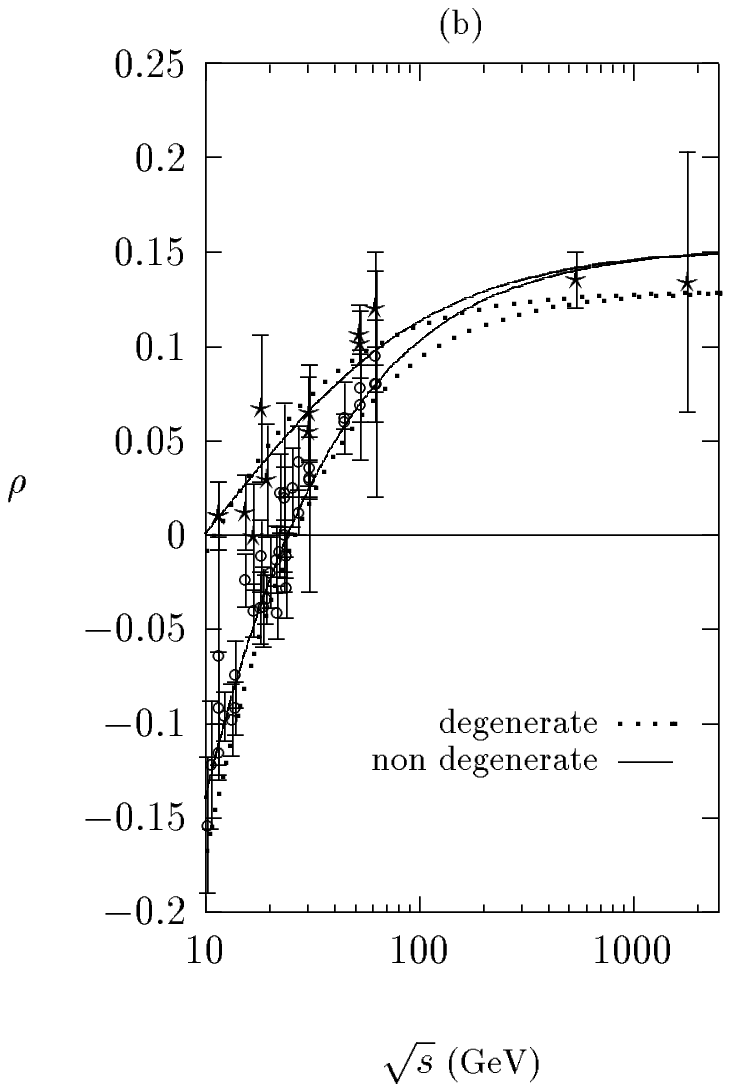,bbllx=2.5cm,bblly=12cm,bburx=12cm,bbury=23cm,width=6cm}
}
\begin{quote}
{\small Figure 2:  Best fits to $2\sigma$ filtered data. The dotted lines
correspond to
the original DL model given in Eq.~(\protect{\ref{amplitude}}), whereas the
plain ones correspond
to a model where the degeneracy of the lower trajectories is lifted, as in
Eq.~(\protect{\ref{2traj}}). The data points are the PDG data filtered at
the $2\sigma$ level.
}
\label{bestfits}
\end{quote}
\begin{center}
{\small \begin{tabular}{|c|c|c|c|}\hline
parameter& all data &filtered data ($2\sigma$)&filtered data ($1\sigma$)\\
\hline
$\chi^2	$		& 505.4&119.6&57.6\\
$\chi^2$ per d.o.f.	& 2.99		& 0.77&0.47\\\hline
pomeron intercept--1	&$0.0990^{+0.0099}_{-0.0088}$&
$0.0964^{+0.0115}_{-0.0091}$&$0.095^{-0.013}_{+0.010}$\\
pomeron coupling (mb)	&$17.5^{+1.9}_{-2.0}$&
$18.0^{+2.0}_{-2.2}$&$18.2^{+2.3}_{-2.6}$\\\hline
$\rho/\omega$ intercept--1	&
$-0.494^{+0.056}_{-0.066}$&$-0.498^{+0.057}_{-0.067}$
&$-0.510^{+0.064}_{-0.077}$\\
$\rho$/$\omega$ coupling $C_-(pp)$ (mb)& $-24.0^{+6.8}_{-10.9}$ &
$-26.5^{+7.7}_{-12.5}$&$-28.2^{+8.9}_{-15.4}$\\\hline
$a/f$ intercept--1		&$-0.312^{+0.051}_{-0.052}$&$-0.315\pm 0.058$&$-0.324\pm
0.066$\\
$a/f$ coupling (mb)		&
$56.8^{+8.1}_{-6.7}$&$54.9^{+9.0}_{-7.2}$&$56.2^{+9.9}_{-7.8}$\\\hline
$\sigma_{tot}(1.8$ TeV$)$ (mb)	&
$77.6^{+2.5}_{-2.7}$&$76.8^{+2.9}_{-2.7}$&$76.4^{+3.4}_{-3.1}$\\\hline
$\sigma_{tot}(10$ TeV$)$ (mb)	&
$108.4^{+7.0}_{-6.7}$&$106.4^{+7.9}_{-6.7}$&$105.4^{+9.2}_{-7.5}$\\\hline
$\sigma_{tot}(14$ TeV$)$ (mb)	&
$115.8^{+8.3}_{-7.7}$&$113.5^{+9.3}_{-7.8}$&$112.3^{+10.8}_{-8.6}$\\\hline
\end{tabular}}~\\
\end{center}\begin{quote}
{\small Table 4: Simple pole fit to total $pp$ and $\bar p p$ cross sections,
and to the $\rho$
parameter, with non-degenerate $C=+1$ and $C=-1$ meson exchanges.}
\end{quote}
\begin{center}
{\small \begin{tabular}{|c|c|c|c|}\hline
parameter& all data &filtered data ($2\sigma$)&filtered data ($1\sigma$)\\
\hline
$\chi^2	$		& 197.9&107.7&56.3\\
$\chi^2$ per d.o.f.	& 1.39		& 0.82&0.52\\\hline
pomeron intercept--1	&$0.0955^{+0.0097}_{-0.0083}$&
$0.0940^{+0.0092}_{-0.0079}$&$0.095^{+0.013}_{-0.010}$\\
pomeron coupling (mb)	&$18.4^{+1.8}_{-2.0}$&
$18.8^{+1.7}_{-2.0}$&$18.5^{+2.1}_{-2.6}$\\\hline
$\rho/\omega$ intercept--1	&
$-0.535^{+0.051}_{-0.059}$&$-0.518^{+0.050}_{-0.058}$
&$-0.540^{+0.059}_{-0.067}$\\
$\rho$/$\omega$ coupling $C_-(pp)$ (mb)& $-31.6^{+7.6}_{-11.5}$ &
$-28.9^{+6.8}_{-10.4}$&$-32.5^{+8.8}_{-13.9}$\\\hline
$a/f$ intercept--1
&$-0.338^{+0.054}_{-0.055}$&$-0.355^{+0.056}_{-0.057}$
&$-0.346^{+0.067}_{-0.066}$\\
$a/f$ coupling (mb)		&
$58.8^{+8.7}_{-6.8}$&$61.5^{+9.8}_{-7.7}$&$60.4^{+10.5}_{-7.9}$\\\hline
$\sigma_{tot}(1.8$ TeV$)$ (mb)	& $77.3^{+2.6}_{-2.7}$&$77.2\pm
2.6$&$77.2^{+3.6}_{-3.3}$\\\hline
$\sigma_{tot}(10$ TeV$)$ (mb) 	&
$106.8^{+7.1}_{-6.3}$&$106.3^{+6.8}_{-6.2}$&$106.5^{+9.7}_{-7.8}$\\\hline
$\sigma_{tot}(14$ TeV$)$ (mb)	&
$113.9^{+8.3}_{-7.4}$&$113.2^{+8.0}_{-7.1}$&$113.5^{+11.4}_{-9.0}$\\\hline
\end{tabular}}~\\
\end{center}\begin{quote}
{\small Table 5: Simple pole fit to total $pp$ and $\bar p p$ cross sections,
and to the $\rho$
parameter, with non-degenerate $C=+1$ and $C=-1$ meson exchanges, and using the
alternative data set of Ref.~\cite{Kangdata}.}
\end{quote}
%
%In order to better understand the treatment of the errors, we give in Table~5
%the result of a fit to the data of Ref.~\cite{Kangdata}. We see that the %results are very stable, especially those for the pomeron intercept, %independently of the data filtering. 
Our best estimate for the pomeron
intercept is then:
\begin{equation}
\alpha_P=1.0964^{+0.0115}_{-0.0091}
\label{result}
\end{equation}
based on the $2\sigma$-filtered PDG data, in the non-degenerate case.

At this point, the only additional piece of data might be the direct
observation
of the pomeron, {\it i.e.} of a $2^{++}$ glueball. Using the observed X(1900)
mass as confirmed by the WA91 collaboration \cite{WA}, $1918\pm12$ MeV, 
for a $I^GJ^{PC}=0^+2^{++}$ state, $f_2(1900)$, and using
$\alpha'=0.250$ GeV$^{-2}$
\cite{DL}, we obtain
$\alpha_P=1.0803\pm 0.012$.
This is the value of the intercept for 1-pomeron exchange. The intercepts that
we obtained in Tables~2,~3,~4 and 5 from scattering data
cannot be directly compared
with this value, as they include the effect of multiple exchanges,
of pomerons and reggeons. But the values we have derived are certainly
compatible with the WA91 measurement. Note however that 
%the conversion of the
%glueball
%ass into a pomeron intercept relies heavily on the value of the pomeron slope.
an intercept of 1.094 would be in perfect agreement with the WA91 observation
for a slope $\alpha'=0.246$ GeV$^{-2}$. Hence it would be dangerous to mix this
piece of information with the $t$-channel information.

What constraints can we place on physics beyond one-pomeron exchange?
Using $2\sigma$-filtered data, we obtain an upper bound on
the ratio of the 2-pomeron coupling to that of the pomeron to be $4.7 \%$
at 70\% C.L., and including
%\begin{equation}
%{\rm |2-pomeron\ coupling|\over 1-pomeron\ coupling}<4.7\% \ \ \ \ (70\% C.L.)
%\end{equation} 
such a contribution would bring the best value for the intercept of
the 1-pomeron exchange term to $1.126^{+0.051}_{-0.082}$.
Also at the 70\% C.L.,
the ratio of the coupling of an odderon \cite{Odd} to that of a pomeron is smaller
than 0.1\% (the best odderon intercept would then be 1.105 and the pomeron
intercept become 1.099). This would correspond to 0.08 mb at the Tevatron.
As for the ``hard pomeron'', there is no trace of it in the data.
Constraining
its intercept to being larger than 1.3 leads to an upper bound on the ratio
of its coupling
to that of the pomeron of 0.9\% (the soft pomeron intercept then becomes
1.065). This would correspond to a maximum hard contribution of
 19 mb at the Tevatron. Consequently, the simple pole model fits to total cross sections are very successful.

We show in Fig.~3 the results obtained in this
paper together with other estimates present in the literature.
All the
points from this work  have an acceptable $\chi^2$, and the main difference
between them is either the filtering of data or the physics of lower
trajectories. Since all these estimates are acceptable, we conclude that
the pomeron intercepts as high as 1.11, and as low as 1.07, are possible.
When comparing with
other works in the literature, we have explained that the use of a small
energy cutoff leads to smaller intercepts, and reflects
the fact that sub-leading meson trajectories are to be included. Note however
that the original DL fit \cite{DLfit} used the same cutoff as ours,
but used a different definition of $\chi^2$ \cite{peter}. 

%{}~\\
\centerline{
\psfig{figure=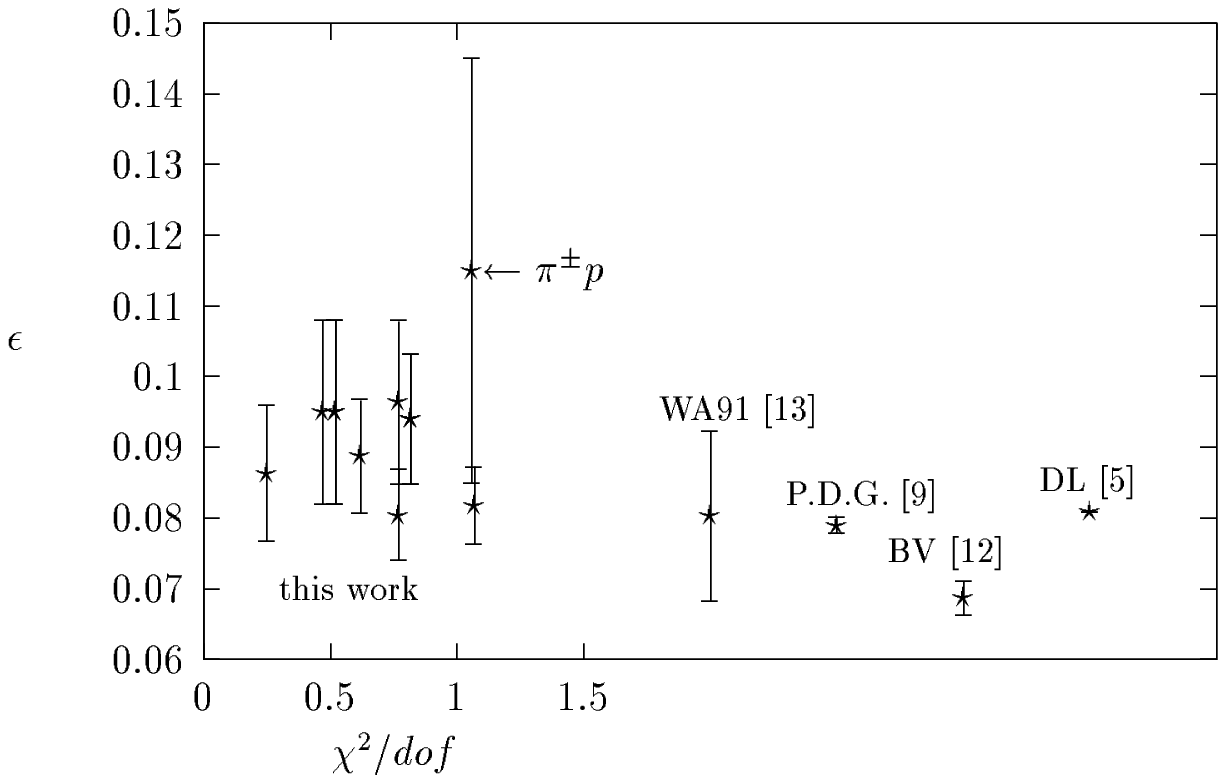,bbllx=2.5cm,bblly=15cm,bburx=17cm,bbury=23cm,width=12cm}
}
\begin{quote}
{\small Figure 3: Our results for the pomeron intercept, compared with others
in the literature. The values of the $\chi^2/d.o.f.$ are indicated for the points
of this work only.}
\end{quote}

Our errors are much larger than those of other estimates because we fully
take into account the correlation of the various parameters, and because
our statistical analysis of the data is much more careful than previous ones.
Though these results depend only on $pp$ and $\bar p p$ data, we have
argued that little could be learned from other hadronic reactions, given
that they are measured at low energy. In particular, we pointed out that
our fit to total cross sections, as shown in Fig.~2, 
is indistinguishable from the DL fit
for $\sqrt{s}<300$ GeV, and hence the parametrisation we propose is expected to fit well the total $\gamma p$ cross sections, as well as the $\pi p$ and $K p$ data. 

In fact the preliminary results of the fit to all $pp$, $\bar p p$, 
$\pi ^{\pm} p$, $K^{\pm} p$, $\gamma\gamma$and $\gamma p$ total cross sections,
for $\sqrt s _{min}$ = 10 GeV gives $\alpha _P = 1.094\pm 0.008$ with $\chi^2 $/d.o.f. =1.62 
but inclusion of the real parts for $pp$, $\bar p p$, $\pi ^{\pm} p$ and $K^{\pm} p$ not only improves the $\chi^2$ to be 1.38 per d.o.f but also gives a very stable central value of the pomeron intercept $1.096\pm 0.006$ \cite{CKK1}. The particle data group finds also the same results for the same data sets of total cross sections and real parts \cite{PDG97}. Also we  have begun to analyse the behavior
of the proton structure function $F_2$ with our best fit parameters \cite{CKK2}.
The preliminary results indicate that we can get a reasonable fits to the ZEUS BPC data up to $q^2 \approx 0.65$ GeV$^2$ beyond which one may have to bring in muti-pomeron effect through unitarization as a larger effective intercept is
needed to fit the data at large $Q^2$\cite{DGLAP}.

\end{document}